# Heisenberg exchange enhancement by orbital relaxation in cuprate compounds

A.B. van Oosten, R. Broer and W.C. Nieuwpoort

Laboratory of Chemical Physics and Materials Science Centre, University of Groningen, Nijenborgh 4, 9747 AG Groningen, The Netherlands

We calculate the Heisenberg exchange J in the quasi-2D antiferromagnetic cuprates $La_2CuO_4$, $YBa_2Cu_3O_6$, $Nd_2CuO_4$ and $Sr_2CuO_2Cl_2$. We apply all-electron (MC)SCF and non-orthogonal CI calculations to $[Cu_2O_{11}]^{18-}$, $[Cu_2O_9]^{14-}$, $[Cu_2O_7]^{10-}$ and $[Cu_2O_7Cl_4]^{14-}$ clusters in a model charge embedding. The (MC)SCF triplet and singlet ground states are well characterized by $Cu^{2+}$ ($d_{x^2-y^2}$) and $O^{2-}$. The antiferromagnetic exchange is strongly enhanced by admixing relaxed (MC)SCF triplet and singlet excited states, in which a single electron is transferred from the central O ion to Cu. We ascribe this effect to orbital relaxation in the charge transfer component of the wave function. Close agreement with experiment is obtained.

Cuprate compounds have been the subject of a massive research effort since the discovery of high Tc superconductivity [1-3]. Nevertheless we are still far from a complete understanding of their fascinating properties. The cuprates consist of $[CuO_2]^{2-}$ planar structural units, which are separated and, to a good approximation, electronically and magnetically insulated from each other by layers of counter-ions. Each Cu site has one unpaired electron with nearly pure Cu-d character [4], which is localised because of Mott-Hubbard correlation. A further complication is that the band gaps, which are typically 1.5 to 2 eV wide, are of the charge transfer type in stead of the Mott-Hubbard type. The magnetic behaviour can be described with a Heisenberg hamiltonian

$$H = -J \Sigma_{\langle ij \rangle} \vec{S}_i \cdot \vec{S}_j \qquad (1)$$

where the summation runs over nearest neighbour pairs. Cuprates have negative J values and display long range two-dimensional antiferromagnetic order [5] in the absence of doping or excitation. The antiferromagnetism is qualitatively explained by the superexchange mechanism [6-8], which in the present case involves electron transfer between Cu via one bridging O. Conventional LDA based band structure approaches fail to account for the strong correlation between the Cu holes and predict a half-filled Cu-O hybridised band[9], whereas one expects a filled lower and an empty upper Hubbard band, separated by approximately $U_d$, and a filled oxygen band situated above the former.

The electronic structure of cuprates is even more complicated when additional holes are present. Such holes reside preferably on in-plane O sites [10] and display strong local relaxation effects [11-14]. Also in O 2p to Cu 3d charge transfer excited states relaxation effects have been shown to be very important[16]. Because the superexchange mechanism involves charge transfer excitations, such effects have to be taken into account in the calculation of J. Correlation and local relaxation can be described with *ab initio* quantum chemical cluster methods. An early GVB study of $La_2CuO_4$ [15] yielded a value of J=-35 meV. A standard SD-CI calculation gives J=-69 meV, or J=-83 meV if the Davidson estimate for the contributions of quadruples is included [16]. In a recent ab initio cluster model study by Casanovas et al. [17] a best calculated value of 98 meV was obtained for J. The experimental values are J=-128 meV [18,19] from Raman measurements and J=-134 meV [20] from neutron scattering. The above results demonstrate the capability of the *ab initio* cluster approach to properly predict the antiferromagnetic ground state of La2CuO4 and related materials. Casanovas et al. [17] have suggested that the remaining difference of about 20 % with the experimental value is due to the unability of two-center models to give a quantitative description of J. It is shown below, however, that a more complete inclusion of local relaxation and correlation effects in the charge transfer state does give rise to a correct prediction of J within the two-center model.

In a previous paper [14] we obtained J=-120 meV for $La_2CuO_4$. Here we present results on four antiferromagnetic cuprate compounds. Three of the cuprates we study, $La_2CuO_4$ [1], $YBa_2Cu_3O_6$ [2] and $Nd_2CuO_4$ [3], are superconductor parent compounds. The fourth cuprate, $Sr_2CuO_2Cl_2$, is closely related to superconductor compounds. We selected these compounds because they have been studied by Raman [18,19,21,22] and neutron scattering [5,20,23,24]. From the data accurate values for J have been derived for $La_2CuO_4$, $YBa_2Cu_3O_6$ and $Nd_2CuO_4$, which are listed in table 2. Also listed is a literature estimate for $Sr_2CuO_2Cl_2$ that we will discuss below.

We perform (MC)SCF and non-orthogonal CI calculations on the basic planar $Cu_2O_7$ cluster occurring in all four compounds, extended with the first out-of-plane O or Cl neighbours of the Cu ions. For $La_2CuO_4$, $YBa_2Cu_3O_6$, $Nd_2CuO_4$ and $Sr_2CuO_2Cl_2$ we employ $Cu_2O_{11}$ ($D_{2h}$), $Cu_2O_9$ ($C_{2v}$), $Cu_2O_7$ ($D_{2h}$) or $Cu_2O_7Cl_4$ ($D_{2h}$) clusters, respectively. As an example, the $Cu_2O_{11}$ cluster as used for $La_2CuO_4$ is shown in figure 1. Structural data were taken from references [25-28], respectively. The clusters are embedded in a point charge environment that accurately represents the Coulomb potential inside the cluster region of a surrounding infinite lattice of formal ionic charges. The use of the simple point charges at nearest neigbour positions may lead to spurious occupation of diffuse orbitals, because these orbitals feel the strong attractive potential of the $Cu^{2+}$ and $La^{3+}$ point charges, without being repelled by the ion core [13, 16]. Therefore in the present study the potential due to the ions at the nearest neighbour positions to the cluster, is modified to be flat inside a small sphere with an ionic radius around these point charges. These modified potentials improve the stability against variation of the most diffuse components of the basisset, whereas in the cluster region the potential due to the modified charges is identical to that of the bare point charges. The complete specification of the background potential is available from the authors on request.

As a first step we perform all-electron (MC)SCF calculations of the lowest singlet and triplet states. We introduce local exchange and correlation effects on the bridging oxygen atom through admixture of excited (MC)SCF states, that differ from the MCSCF ground states by an O -> Cu electron excitation. This is an example of non-orthogonal CI (NOCI) [29]. NOCI involves the computation of Hamilton and overlap matrix elements between determinants constructed from non-orthogonal orbital sets [30]. It has the advantage over conventional CI methods that it leads to a short, physically transparent wave function and that it is free of size consistency errors. We use contracted Gaussian basis sets: for Cu we adopt the Wachters (14s,9p,5d) basis set optimised for the $^2$D state [31], augmented with the Hay diffuse d-function [32] and two diffuse p-functions (exponents 0.31 and 0.12). The final (14s,11p,6d) basis set is contracted to (8s,6p,3d). For O a (9s,5p) Huzinaga basis set, contracted to (3s,2p) [33], was extended with a diffuse p-function (exponent 0.11), giving a final (3s,3p) contracted oxygen basis. For Cl we use the (12s,8p) Huzinaga basis set contracted to (6s,4p) [33].

In an SCF wavefunction for the lowest triplet state of the undoped model clusters two open shell orbitals occur, denoted by $d_g$ and $d_u$, respectively. This wavefunction can be written as

$$\Psi_t = |\,\sigma\bar{\sigma}d_g d_u\,|\,. \tag{2a}$$

The subscripts g and u denote even and odd symmetry under inversion. For the $D_{2h}$ clusters that we use to describe La$_2$CuO$_4$, Nd$_2$CuO$_4$ and Sr$_2$CuO$_2$Cl$_2$, $d_g$ and $d_u$ transform as $a_g$ and $b_{3u}$, respectively. For YBa$_2$Cu$_3$O$_6$, where C$_{2v}$ symmetry applies, $d_g$ and $d_u$ transform as $a_1$ and $b_1$. In (2a) $\sigma$ denotes the O(2p$_\sigma$) orbital at the bridging oxygen. It has the same symmetry as $d_u$. For clarity all other closed shell orbitals are suppressed in the notation. The calculated triplet SCF ground states for each compound are very well characterised by Cu$^{+2}$ (3d$^9$) and O$^{-2}$ (2p$^6$), respectively. As an example, table 1 gives the open shell Mulliken populations of the triplet ground state of the $Cu_2O_{11}$ cluster describing $La_2CuO_4$. The Cu holes are seen to have almost pure $3d(x^2-y^2)$ character, in agreement with experiment [4].

An alternative way to write the triplet SCF wavefunction is in terms of symmetry equivalent, mutually orthogonal open shell orbitals $d_1$ and $d_2$:

$$\Psi_t = |\sigma\bar{\sigma}d_1d_2|. \tag{2b}$$

in which $d_1$ and $d_2$ are defined by

$$d_1 = \frac{1}{\sqrt{2}}(d_g + d_u),$$
$$d_2 = \frac{1}{\sqrt{2}}(d_g - d_u), \tag{3}$$

The singlet corresponding to the triplet (2) is

$$\Psi_s = \frac{1}{\sqrt{2}}(|\sigma\bar{\sigma}d_1\bar{d}_2| - |\sigma\bar{\sigma}\bar{d}_1d_2|) \tag{4}$$

Within this description we obtain the direct exchange contribution. As shown in table 1, the direct exchange ranges between J=+13 meV and J=+15 meV, if the triplet orbitals are used to describe the singlets. If the orbitals are separately optimised for $\Psi_s$ and $\Psi_t$, we obtain J≈+6 meV for all four compounds.

Let us now, in (2b) and (4), replace the open shell orbitals $d_1$ and $d_2$ by normalised, mutually nonorthogonal orbitals

$$d'_1 = \frac{d_1 + \lambda d_2}{\sqrt{1+\lambda^2}},$$

$$d'_2 = \frac{d_2 + \lambda d_1}{\sqrt{1+\lambda^2}}, \tag{5}$$

with overlap $S = \langle d'_1 | d'_2 \rangle = \frac{2\lambda}{\sqrt{(1+\lambda^2)}}$. The singlet wave function can then be written as

$$\Psi_s = (2+2S^2)^{-1/2} \,|\, \sigma\bar{\sigma} \,( d'_1 \bar{d}'_2 - \bar{d}'_1 d'_2 ) \,|$$

$$= (2+2S^2)^{-1/2} \,\{|\, \sigma\bar{\sigma} \,( d_1 \bar{d}_2 - \bar{d}_1 d_2 ) + S \,|\, \sigma\bar{\sigma} \,( d_1 \bar{d}_1 + d_2 \bar{d}_2 ) \,|\,\},$$

$$= (2+2S^2)^{-1/2} \,\{|\, (1+S) \,\sigma\bar{\sigma} d_g \bar{d}_g - (1-S) \,\sigma\bar{\sigma} d_u \bar{d}_u \,|\,\}, \qquad (6)$$

For the singlet wave function the substitution (5) introduces an additional variational parameter $\lambda$ or, equivalently, S. Contrarily, for the triplet (2) the substitution (5) has no effect and S can be set to zero without loss of variational freedom. Note also that the wave function (6) is identical to the CASSCF wave function with $d_g$ and $d_u$ in the active space, provided that $\lambda$ as well as the orbitals are optimized. The wave functions (2) and (6) form an appropriate starting point of a balanced calculation of the singlet-triplet splitting [34]. The splittings correspond to the Anderson superexchange [7]. The singlet states (6) with optimized orbitals all have S≈0.04 and their energies are 20 to 30 meV below the triplets. These values should be compared to the value of J=-35 meV obtained by Guo et al. [15] for $La_2CuO_4$. These authors used an equivalent wave function but slightly different basis sets and a point charge embedding.

Anderson superexchange is due to charge transfer excitations of the type $d_1)^1 d_2)^1 \to d_1)^2 d_2)^0 + d_1)^0 d_2)^2$, which can only occur for the singlet. Geertsma [8] has discussed charge transfer excitations of a different type, namely $d_1)^1 \sigma)^2 d_2)^1 \to d_1)^2 \sigma)^0 d_2)^2$. These can be included by extending the active space to $d_g$, $d_u$ and $\sigma$. The corresponding CASSCF wave function is

$$\Psi_s = c_1 \,|\, \sigma\bar{\sigma} d_g \bar{d}_g \,| + c_2 \,|\, \sigma\bar{\sigma} d_u \bar{d}_u \,| + c_3 \,|\, d_g \bar{d}_g d_u \bar{d}_u \,| + c_4 \,|\, ( d_g \bar{d}_g \,(\sigma \bar{d}_u - \bar{\sigma} d_u ) \,|\,. \qquad (7)$$

This was rewritten as a linear combination of closed shell determinants and treated in the Hartree-Fock-Roothaan scheme. We find that the additional degree of freedom does not lower the energy of the singlet state.

As a next step we admix to $\Psi_t$ and $\Psi_s$ relaxed charge transfer excitations of the form $d_1)^1 \sigma)^2 d_2)^1 \to d_1)^1 \sigma)^1 d_2)^2 \pm d_1)^2 \sigma)^1 d_2)^1$. Admixture of *unrelaxed* excitations of this type has no effect [35], but we find that this is quite different for the relaxed charge transfer excitations. The excited states of $^3B_{3u}$ ($^3B_1$) and $^1A_g$ ($^1A_1$) symmetry are

$$\Psi_t^* = |\, d_u \bar{d}_u d_g \sigma\, |, \qquad (8)$$

$$\Psi_s^* = \frac{1}{\sqrt{2}} (\,|\, d_g \bar{d}_g d_u \bar{\sigma}\, | - |\, d_g \bar{d}_g \bar{d}_u \sigma\, |\,). \qquad (9)$$

The orbitals of the triplet state $\Psi_t^*$ could not be optimised, because the SCF process converged to the triplet ground state, $\Psi_t$. We therefore use the orbitals of the corresponding singlet wave function, $\Psi_s^{**} = \frac{1}{\sqrt{2}} (\,|\, d_u \bar{d}_u d_g \bar{\sigma}\, | - |\, d_u \bar{d}_u \bar{d}_g \sigma\, |\,)$, which has $^1B_{3u}$ ($^1B_1$) symmetry. This is a reasonable approximation, because the exchange integral between the $a_g$ and the $b_{3u}$ orbital for this wave function is only 0.20 eV, to be compared to the energy separation of about 10 eV between $\Psi_t$ and $\Psi_t^*$.

In the excited singlet state $\Psi_s^*$ (10) again an overlap between the open shell orbitals is allowed, just as in the case of $\Psi_s$. Moreover, the open shell orbitals are allowed to mix because they have the same symmetry. We therefore consider normalised, mutually non-orthogonal orbitals of the form

$$d'_u = \frac{d_u + \lambda_1 \sigma}{\sqrt{1+\lambda_1^2}},$$

$$\sigma' = \frac{\sigma + \lambda_2 d_u}{\sqrt{1+\lambda_2^2}}. \qquad (10)$$

Their overlap is

$$S^* = \frac{\lambda_1 + \lambda_2}{\sqrt{\{(1+\lambda_1^2)(1+\lambda_2^2)\}}}. \qquad (11)$$

This transformation does not alter $\Psi_t^*$, while $\Psi_s^*$ acquires extra variational freedom. $\Psi_s^*$ may then be written in the form

$$\Psi_s^* = (2+2S^2)^{-1/2} \ \{|\,d_g\bar{d}_g d'_u\bar{\sigma}'\,| - |\,d_g\bar{d}_g \bar{d}'_u\sigma'\,|\} \ . \tag{12a}$$

Equivalently, $\Psi_s^*$ can be written in terms of mutually orthogonal open shell orbitals, $d_u$ and $\sigma$, as a CASSCF wave function with $d_u$ and $\sigma$ in the active space

$$\Psi_s^* = c_1 \,(\,|\,d_g\bar{d}_g d_u\bar{\sigma}\,| - |\,d_g\bar{d}_g \bar{d}_u\sigma\,|\,) + c_2\,|\,d_g\bar{d}_g d_u\bar{d}_u\,| + c_3\,|\,d_g\bar{d}_g \sigma\bar{\sigma}\,|. \tag{12b}$$

A third equivalent form, which was actually employed in the calculations, is a linear combination of two closed shell configurations

$$\Psi_s^* = \frac{(1+S^*)\,|\,d_g\bar{d}_g u_1\bar{u}_1\,| - (1-S^*)\,|\,d_g\bar{d}_g u_2\bar{u}_2\,|}{\sqrt{(2+2S^{*2})}} \ . \tag{12c}$$

Here new orbitals $u_1$ and $u_2$ are introduced that are related to $d_u$ and $\sigma$ by an orthogonal transformation

$$\sigma = \cos\alpha\ u_1 + \sin\alpha\ u_2\ ,$$

$$d_u = \sin\alpha\ u_1 - \cos\alpha\ u_2\ ,$$

$$\cos\alpha = \sqrt{\left\{\frac{1}{2} + \frac{1}{2\sqrt{1+p^2}}\right\}}\ ,$$

$$p = \frac{1+\lambda_1\lambda_2}{\lambda_1-\lambda_2}\ . \tag{13}$$

A Mulliken population analysis of the optimized $u_1$ and $u_2$ orbitals of equation (12c), as obtained for the $Cu_2O_{11}$ cluster representing $La_2CuO_4$ is given in table 1. We find $\lambda_1 \approx \lambda_2$ and $S^* \approx 0.66$ and similar results for the other three cuprates. These results indicate a tendency in the excited singlet state towards covalent bond formation between the bridging O and the remaining $Cu^{2+}$ neighbour, which is enhanced by orbital relaxation. These effects are absent in $\Psi_t^*$.

A NOCI between the wave functions (2) and (8) lowers the energy of the triplet state by an energy varying from 15-35 meV for the four cuprate compounds. The singlet NOCI is performed in three steps using progressively more accurate wave functions for $\Psi_s^*$. First we mix (6) with (9), which has orthogonal orbitals. We find that the singlet energy is lowered by the same amount as the triplet, so that J is unaltered. A slightly larger energy lowering of singlet, leading to a slightly larger singlet-triplet splitting, occurs if we admix (12) to (6). In the final NOCI calculation, we allow S in (6) and the expansion coefficients in (12b) to be reoptimised. This amounts to reoptimising $\lambda$ in (5), as well as $\lambda_1$ and $\lambda_2$ in (11). Since the triplet wave function is invariant under changes in $\lambda$, $\lambda_1$ and $\lambda_2$, this procedure achieves the correct variational balance. In this final calculation a substantial differential effect is found, which accounts for more than half of the calculated exchange splitting.

The results of the different stages of the calculation for the different compounds are listed and compared to the experimental values in table 2. The discrepancy between the theoretical and the literature value of J for $Sr_2CuO_2Cl_2$ deserves further comment. The calculation gives a value close to that of $Nd_2CuO_4$. As the values of d and $\Delta V_m$ for $Sr_2CuO_2Cl_2$ and $Nd_2CuO_4$ are quite close, this result is reasonable. Our result also agrees with Raman spectra [21]. The spectrum of $Sr_2CuO_2Cl_2$ coincides with that of $Nd_2CuO_4$ but is downshifted with respect to $La_2CuO_4$. This leads us to believe that an analysis of $Sr_2CuO_2Cl_2$ data with the approach of reference [18] should improve the agreement with our calculated result.

From our results we conclude that the Heisenberg exchange in cuprate compounds is strongly enhanced by orbital relaxation. This follows also from the fact that a CASSCF with $d_u$, $d_g$ and $\sigma$ in the active space, as in (7), is insufficient, while admixture of a relaxed charge transfer excitation has a large effect.

We find that $\Psi_s^*$ as given by (12) has a large overlap with $\Psi_s$, which is the lowest root CASSCF wave function with $d_u$, $d_g$ and $\sigma$ in its active space. $\Psi_s^*$, however, is the lowest root CASSCF wave function with the smaller ($d_u$, $\sigma$) active space. It is plausible that reoptimisation of $\lambda$, $\lambda_1$ and $\lambda_2$ allows $\Psi_s^*$, after orthogonalisation to $\Psi_s$, to approximate the second root CASSCF wave function in the ($d_u$, $d_g$, $\sigma$) active space. This suggests two alternative approaches for future research: 1) Calculation of the first and the second root CASSCF wave functions in the ($d_u$, $d_g$, $\sigma$) active space followed by NOCI between these, or 2) Extension of the active space to include an extra set of valence orbitals to describe the orbital relaxation that occurs when a hole moves to the bridging O.

We have repeated the $La_2CuO_4$ calculation on a $Cu_2O_7$ cluster and obtained 110 meV, in good agreement with the 120 meV obtained for the $Cu_2O_{11}$ cluster. This gives confidence that our calculation is reasonably stable against cluster size effects.

In summary, we have shown that orbital relaxation in the charge transfer component of the wave function is responsible for the large Heisenberg exchange observed in cuprate compounds.

We gratefully acknowledge the financial support of the Netherlands Foundation of Fundamental Research on Matter (FOM). This research was sponsored by the Netherlands National Computer Facilities Foundation (NCF) for the use of supercomputer facilities, with financial support from the Netherlands Organization for Scientific Research (NWO). We also thank Dr. W. Geertsma for many valuable discussions.

TABLES

| state | orbital symmetry | orbital occupation | Cu 3d ($3z^2 - r^2$) | Cu 3d ($x^2 - y^2$) | O 2p (x) |
|---|---|---|---|---|---|
| $\Psi_s$ | $a_g$ | 0.54 | 0.000 | 0.947 | 0 |
|  | $b_{3u}$ | 0.46 | 0.000 | 0.913 | 0.030 |
| $\Psi_t$ | $a_g$ | 1.00 | 0.000 | 0.945 | 0 |
|  | $b_{3u}$ | 1.00 | 0.000 | 0.917 | 0.028 |
| $\Psi_s^*$ | $b_{3u}$ | 0.96 | 0.004 | 0.352 | 0.589 |
|  | $b_{3u}$ | 0.04 | 0.009 | 0.461 | 0.428 |
| $\Psi_t^*$ | $a_g$ | 1.00 | 0.363 | 0.487 | 0 |
|  | $b_{3u}$ | 1.00 | 0.034 | 0.057 | 0.847 |

Table 1. Occupation and Mulliken population of the open shell orbitals of $[Cu_2O_{11}]^{18-}$ ground and excited states ($La_2CuO_4$).

|  | a | b | c | d | e |
|---|---|---|---|---|---|
| $La_2CuO_4$ | 15 | 9 | -30 | -120 | -128[1] |
| $YBa_2Cu_3O_6$ | 13 | 6 | -22 | -98 | -98[1] |
| $Nd_2CuO_4$ | 13 | 6 | -23 | -102 | -108[1] |
| $Sr_2CuO_2Cl_2$ | 13 | 6 | -22 | -106 | -125[2] |
|  |  |  |  |  | ~-108[3] |

Table 2. Calculated J: a) Singlet constructed from triplet orbitals (eq. 2); b) Relaxed, orthogonal singlet (4); c) Relaxed, non-orthogonal singlet (6); d) Non-orthogonal CI involving (6) and (12b), see text; e) Experiment. [1]Ref. [18]; [2]Ref. [23]; [3]Ref. [21] and see text.

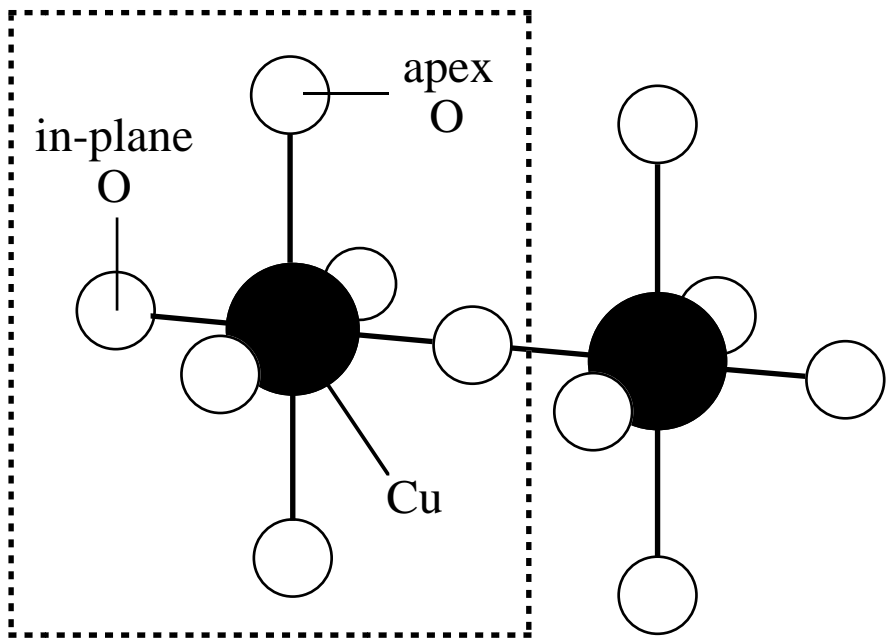

Figure 1. The cluster model $Cu_2O_{11}$.

REFERENCES

[1]  J.G. Bednorz and K.A. Müller, Z. Phys. B 64 (1986) 189.

[2]  M.K. Wu, J.R. Ashburn, C.J. Torng, P.H. Hor, R.L. Meng, L. Gao, Z.J. Huang, Y.Q. Wang and C.S. Chu, Phys. Rev. Lett. 58 (1987) 908.

[3]  Y. Tokura, H. Takagi and S. Uchida, Nature 337 (1989) 345.

[4]  A. Fujimori, E. Takayama-Muromachi, Y. Uchida and B. Okai, Physical Review B 35 (1987) 8814.

[5]  D. Vaknin, S.K. Sinha, D.E. Moncton, D.C. Johnston, J.M. Newsam, C.R. Safinya and H.E. King, Phys. Rev. Lett. 58 (1987) 2802.

[6]  H.A. Kramers, Physica 1 (1934) 182.

[7]  P.W. Anderson, Phys. Rev. 115 (1959) 2.

[8]  W. Geertsma, Physica B 164 (1990) 241.

[9]  L.F. Mattheis, Phys. Rev. Lett. 58 (1987) 1028.

[10] N. Nücker, J. Fink, J.C. Fuggle, P.J. Durham and W.M. Temmerman, Physical Review B 37 (1988) 5158.

[11] H. Burghgraef, Department of Chemistry Doctoral Scription, Groningen (The Netherlands): University of Groningen (1990).

[12] A.B. van Oosten, R. Broer, B.T. Thole and W.C. Nieuwpoort, J. Less-Comm. Metals 164-165 (1990) 1514.

[13] R. Broer, A.B. Van Oosten and W.C. Nieuwpoort, Rev. Solid State Sci. 5 (1991) 79.

[14] A.B. van Oosten, R. Broer and W.C. Nieuwpoort, Int. J. Quant. Chem. QCS29 (1995) 241.

[15] Y. Guo, J.-M. Langlois and W.A. Goddard III, Science 239 (1988) 896.

[16] R.L. Martin and P.J. Hay, J. Chem. Phys. 98 (1993) 8691.

[17] J. Casanovas, J. Rubio and F. Illas, Phys. Rev. B 53 (1996) 945.

[18] R.P. Singh, P.A. Fleury, K.B. Lyons and P.E. Sulewski, Phys. Rev. Lett. 62 (1989) 2736.

[19] P.E. Sulewski, P.A. Fleury, K.B. Lyons, S.-W. Cheong and Z. Fisk, Physical Review B 41 (1990) 225.

[20] G. Aeppli, S.M. Hayden, H.A. Mook, Z. Fisk, S.-W. Cheong, D. Rytz, J.P. Remeika, G.P. Espinosa and A.S. Cooper, Phys. Rev. Lett. 62 (1989) 2052.

[21] K.B. Lyons, P.A. Fleury, J.P. Remeika, A.S. Cooper and T.J. Negran, Physical Review B 37 (1988) 2353.

[22] Y. Tokura, S. Koshihara, T. Arima, H. Takagi, S. Ishibashi, T. Ido and S. Uchida, Physical Review B 41 (1989) 11657.

[23] D. Vaknin, S.K. Sinha, C. Stassis, L.L. Miller and D.C. Johnston, Physical Review B 41 (1990) 1926.

[24] M. Greven, R.J. Birgeneau, Y. Endoh, M.A. Kastner, B. Keimer, M. Matsuda, G. Shirane and T.R. Thurston, Phys. Rev. Lett. 72 (1994) 1096.

[25] J.M. Longo and P.M. Raccah, J. Solid State Chem. 6 (1973) 526.

[26] M.F. Garbaukas, R.W. Green, R.H. Ahrendt and J.S. Kasper, Inorg. Chem. 27 (1988) 871.

[27] H. Müller-Buschbaum and W. Wollschläger, Z. Anorg. Allg. Chem. 414 (1975) 76.

[28] L.L. Miller, X.L. Wang, S.X. Wang, C. Stassis, D.C. Johnston, J. Faber Jr. and C.-K. Loong, Physical Review B 41 (1990) 1921.

[29] R. Broer and W.C. Nieuwpoort, Theor. Chim. Acta 73 (1988) 405.

[30] The program GNOME was written for this purpose by R. Broer, J. Th. van Montfort and B. Vunderink.

[31] A.J.H. Wachters, J. Chem. Phys. 52 (1970) 1033.

[32] P.J. Hay, J. Chem. Phys. 66 (1977) 4377.

[33] S. Huzinaga, Department of Chemistry Technical Report, Edmonton (Alberta): University of Alberta (1971); H. Dunning and P.J. Hay, in "Modern Theoretical Chemistry", Vol. 3 (Plenum Press, New York, 1977).

[34] R. Broer and W.J.A. Maaskant, Chem. Phys. 102 (1986) 103.

[35] A.J.H. Wachters, Thesis, University of Groningen (1971).